%% file: paper_03.tex
\documentclass[aps,pra,twocolumn,reprint,superscriptaddress,nofootinbib]{revtex4-2}

\usepackage[svgnames,table,xcdraw]{xcolor}
\usepackage{amsmath}
\usepackage{microtype}
\usepackage{amssymb}
\usepackage{braket}
\usepackage{graphicx}
\usepackage{booktabs}
\usepackage{bm}
\usepackage{dsfont}
\usepackage{hyperref}
\usepackage[caption=false]{subfig}
\usepackage{amsthm}

\usepackage{xfrac}
\usepackage{enumitem}
\usepackage{orcidlink}
\usepackage{appendix}
\usepackage[ruled,vlined]{algorithm2e}

\hypersetup{colorlinks,linkcolor={DarkBlue},citecolor={DarkBlue},urlcolor={DarkBlue}}

\let\originalleft\left
                     \let\originalright\right
\renewcommand{\left}{\mathopen{}\mathclose\bgroup\originalleft}
\renewcommand{\right}{\aftergroup\egroup\originalright}

\usepackage{mathtools}

\DontPrintSemicolon
\SetKwInput{KwInput}{Input}
\SetKwInput{KwOutput}{Output}
\SetKwFor{ForEach}{foreach}{do}{}
\SetKwFor{ForLoop}{for}{do}{}

\input{preamble.tex}

\begin{document}

    \title{%
        \texorpdfstring
        {A Unified Hardware-to-Decoder Architecture for Hybrid Continuous-Variable \\and Discrete-Variable Quantum Error Correction in LiDMaS+}
        {A Unified Hardware-to-Decoder Architecture for Hybrid Continuous-Variable and Discrete-Variable Quantum Error Correction in LiDMaS+}
    }
    \def \affGatech {College of Computing, Georgia Institute of Technology, Atlanta, GA 30332 USA}
    \def \schrosim {Independent Quantum Architecture Researcher \& Software Developer, SchroSIM Quantum Software Project}
    \def \vgg {Volkswagen AG, Berliner Ring 2, Wolfsburg 38440, Germany}
    \def \rwth {Department of Physics, RWTH Aachen, Germany}
    \def \fujf {Department of Computational and Applied Mechanics, Federal University of Juiz de Fora, Juiz de Fora, 36036-900, Brazil}
    \def \lubeck {Institut für Informatik, Fakultät für Mathematik und Informatik, TU Bergakademie Freiberg, Bernhard-von-Cotta-Straße 2, D-09599 Freiberg, Germany}

    \author{Dennis Delali Kwesi Wayo \orcidlink{0000-0001-9980-6247}}
    \affiliation{\affGatech}
    \affiliation{\schrosim}

    \author{Chinonso Onah, \orcidlink{0000-0002-6296-533X}}
    \affiliation{\vgg}
    \affiliation{\rwth}

    \author{Leonardo Goliatt, \,\orcidlink{0000-0002-2844-9470}}
    \affiliation{\fujf}

    \author{Sven Groppe, \,\orcidlink{0000-0001-5196-1117}}
    \affiliation{\lubeck}

    \date{\today}

    \begin{abstract}
        We present an architecture-level hardware-to-logical-to-decoder execution stack for hybrid continuous-variable and discrete-variable quantum error correction in LiDMaS+. Provider-native records are normalized into a single decoder IO contract and replayed under fixed controls across MWPM, UF, BP, and neural-MWPM. In a Xanadu case study using fixture inputs and sampled public datasets, replay integrity was complete: 108/108 fixture and 4000/4000 real-slice request-response lines, with zero request-parse errors, zero response-parse errors, and zero decoder-name mismatches. Under matched inputs, decoder behavior is clearly regime-dependent. For weighted fixture summaries, average flip count was 1.296 (MWPM), 1.296 (UF), 0.667 (BP), and 1.296 (neural-MWPM). For weighted real-data summaries, average flip count was 0.641 (MWPM), 0.741 (UF), 0.318 (BP), and 0.641 (neural-MWPM); corresponding nonempty-flip rates were 0.490, 0.490, 0.318, and 0.490. Across fixture data, BP reduced weighted correction volume by 48.6\% versus MWPM; across real slices, BP reduced weighted correction volume by 50.4\% versus MWPM and 57.1\% versus UF. Quality controls show the central interpretability tradeoff: BP is intervention-conservative but leaves higher residual burden, while MWPM-family decoders intervene more aggressively and clear more syndrome. Warning-no-syndrome rates remained decoder-invariant and dataset-driven (fixture weighted 0.259; real weighted 0.510), confirming preserved sparsity semantics from hardware input to logical correction. Re-running analysis stages reproduced identical SHA-256 artifacts, enabling deterministic study iteration. These results establish a practical benchmarking foundation for photonic GKP-oriented hardware programs where decoder policy must be selected as a function of operating regime.
    \end{abstract}

    \maketitle

    \section{Introduction}
    Quantum error correction (QEC) is entering an architecture phase in which
    hardware data structure, logical encoding choices, and decoder policy must be
    evaluated as one coupled system. Surface-code experiments have established
    strong discrete-variable baselines across multiple platforms
    \cite{acharya2023suppressing,google2025belowthreshold}. In parallel,
    continuous-variable (CV) hardware grounded in GKP encoding
    \cite{gkp2001} is progressing from theory to integrated photonic
    realizations \cite{tzitrin2021static,larsen2025integratedgkp}, with
    complementary below-threshold photonic error-reduction demonstrations
    \cite{somhorst2026photondistillation}. In this setting, decoder benchmarking
    is no longer a purely algorithmic exercise; it is a hardware-facing
    architecture problem.

    \begin{figure*}[t]
        \centering
        \includegraphics[width=1.7\columnwidth]{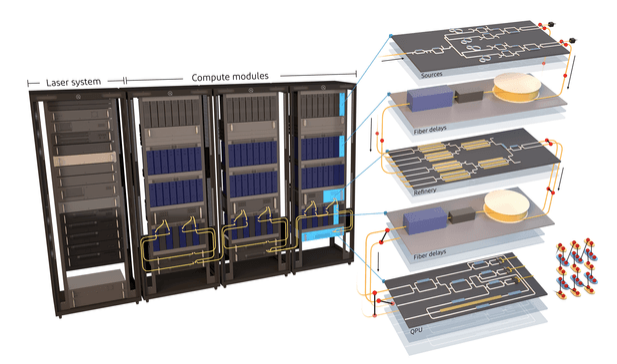}
        \caption{Xanadu photonic experimental stack used to generate the hardware-derived datasets analyzed in this study. Left: rack-level laser, control, and compute modules. Right: conceptual processing chain from optical sources and fiber delays through refinery stages to the QPU/readout layer. This visual documents physical provenance for replay inputs in the LiDMaS+ architecture.}
        \label{fig:xanadu_hardware_photo_placeholder}
    \end{figure*}

    Figure~\ref{fig:xanadu_hardware_photo_placeholder} anchors the physical
    provenance of the data path used in this work. The key challenge is that
    hardware providers expose heterogeneous native outputs (shot records, switch
    traces, count summaries, and matrix structures), while decoder studies need a
    stable logical representation of syndrome events. Without a fixed interface
    between these layers, measured decoder differences can be contaminated by
    ingestion drift rather than true policy behavior.

    This study establishes a unified hardware$\rightarrow$logical$\rightarrow$decoder
    execution architecture in LiDMaS+ \cite{wayo2026lidmas}. Provider-native
    records are mapped to a single line-oriented decoder IO contract; identical
    request streams are then replayed across MWPM, UF, BP, and neural-MWPM under
    fixed controls. The resulting design separates two axes that are usually
    entangled: (i) input-structure effects inherited from hardware data and
    (ii) correction-policy effects introduced by decoder choice.

    This separation is especially important for photonic GKP-oriented programs.
    GKP and hybrid CV-DV stacks expose calibration-sensitive, sparsity-varying
    syndrome traffic that changes with operating regime. A benchmarking method
    that preserves hardware semantics while enforcing decoder comparability is
    therefore directly relevant to deployment decisions, not only to offline
    analysis.

    The decoder literature is broad: MWPM remains a strong topological baseline
    \cite{pymatching2021,fowler2012proofmatching}, union-find variants target
    low overhead \cite{unionfind2017,duclos2010fast}, parallel-window methods
    improve scaling \cite{skoric2023parallel,tan2023scalable}, and neural
    decoders continue to advance throughput and logical suppression in practical
    settings \cite{gu2026scalableneural}. Across these families, fair comparison
    depends on stable input contracts and auditable replay controls. Our
    contribution is to provide that architecture layer and to quantify decoder
    behavior under matched hardware-derived inputs.

    The first case study is Xanadu-oriented and includes fixture conversions plus
    controlled real-data slices. Within this scope, we answer four architecture
    questions:
    \begin{enumerate}[leftmargin=*]
      \item Can heterogeneous provider-native inputs be mapped to one decoder IO
      schema with full parser stability and line-level coverage?
      \item Under fixed requests, do decoders exhibit separable correction
      signatures while input diagnostics remain decoder-invariant?
      \item Can staged analysis be regenerated deterministically at artifact
      level?
      \item Can decoders be swapped as execution engines by configuration, with
      no ingestion rewrites?
    \end{enumerate}

    Beyond interface correctness, the central scientific takeaway is
    regime-dependent decoder behavior. Across sparse and denser request regimes,
    BP is consistently intervention-conservative (lower flip volume) while
    MWPM-family decoders are more intervention-aggressive and clear more
    syndrome. This intervention-versus-residual tradeoff is the key interpretive
    signal for architecture-level decoder selection.

    \subsection{ Novel Contributions}
    This study contributes:
    \begin{itemize}
        \item A unified hardware-to-logical decoder IO contract in LiDMaS+ that
        preserves event, noise, and provenance semantics across heterogeneous
        hardware-derived sources.
        \item A provider-extensible execution architecture that localizes native
        mapping logic while keeping decoder replay logic shared.
        \item A deterministic staged analysis stack that supports auditable
        regeneration of matrix summaries and comparative figures.
        \item Empirical evidence that decoder policy separability is
        regime-dependent under matched inputs, with BP, MWPM/UF, and neural-MWPM
        occupying distinct intervention-residual operating points.
    \end{itemize}

    Together, these contributions position LiDMaS+ as a practical architecture
    substrate for photonic QEC programs: hardware records can be evaluated with
    consistent logical semantics, decoders can be compared under controlled
    conditions, and policy decisions can be tied to operating regime rather than
    tooling variance.

    \section{Unified Hardware-to-Decoder Architecture}
    \label{sec:workflow}
    This section defines the architecture and contract used across providers.

    \subsection{Decoder IO Contract}
    LiDMaS+ uses a line-oriented request/response contract so that
    hardware-derived records and decoder artifacts can be replayed and analyzed in
    matrix form without format-specific branches.
    A request line contains:
    \begin{itemize}
        \item \texttt{code\_id}, \texttt{round\_index}, and \texttt{n\_qubits} for code/round identity,
        \item \texttt{events} as a list of syndrome events (\texttt{index}, \texttt{time\_ns}, \texttt{type}),
        \item \texttt{noise} with scalar priors (\texttt{sigma}, gate/measurement/idle rates, optional loss list),
        \item \texttt{metadata} for provenance (provider, backend, source format, job/shot identifiers).
    \end{itemize}
    A response line contains:
    \begin{itemize}
        \item \texttt{correction} (\texttt{qubit\_flips}, decoder label, optional confidence),
        \item \texttt{diagnostics} (\texttt{sx\_count}, \texttt{sz\_count}, warning/error flags).
    \end{itemize}
    Contract validation in this study checks parseability, request/response
    line-count consistency, decoder-name consistency, and nonfatal diagnostics
    (e.g., \texttt{warning=no\_syndrome\_bits}).

    \subsection{Provider-Oriented Integration Layout}
    The provider contract is intentionally narrow so additional vendors can be
    integrated with the same method:
    \begin{enumerate}[leftmargin=*]
      \item provide a converter from provider raw format to decoder IO request records,
      \item provide a mapping file that binds provider observables to stabilizer events,
      \item provide a data-extraction step that emits request files into the shared layout.
    \end{enumerate}
    This isolates provider-specific ingestion from decoder-specific replay.

    \subsection{Conversion and Replay Stages}
    The execution flow has two core steps. Conversion normalizes heterogeneous
    inputs to a common event stream with consistent noise priors and metadata,
    writing one request line per shot/round. Replay then executes each request
    set across a selected decoder list under matched controls to produce a
    complete dataset-by-decoder matrix.

    \begin{algorithm}[t]
      \caption{Hardware-to-decoder replay and validation pipeline}
      \label{alg:hw_decoder_pipeline}
      \KwInput{Dataset set $\mathcal{D}$, mapping set $\mathcal{M}$, decoder set $\mathcal{K}$, shot cap $S$}
      \KwOutput{Validated replay matrix $\mathbf{R}$ with per-cell diagnostics}
      Initialize empty matrix $\mathbf{R}$\;
      \ForEach{$d \in \mathcal{D}$}{
          raw $\leftarrow$ LoadOrFetchDataset$(d)$\;
          map $\leftarrow$ LoadMapping$(\mathcal{M}[d])$\;
          requests $\leftarrow$ ConvertToDecoderIO$(raw, map, S)$\;
          req $\leftarrow$ AnalyzeRequestLines$(requests)$\;
          ValidateRequestSchema$(requests)$\;
          \ForEach{$k \in \mathcal{K}$}{
              cfg $\leftarrow$ BuildDecoderConfig$(k)$\;
              responses $\leftarrow$ ReplayDecoderIO$(requests, cfg)$\;
              resp $\leftarrow$ AnalyzeResponseLines$(responses, k)$\;
              stats $\leftarrow$ Merge$(req, resp)$\;
              $N_{\mathrm{req}} \leftarrow$ req.\texttt{request\_lines}\;
              $N_{\mathrm{resp}} \leftarrow$ resp.\texttt{response\_lines}\;
              stats.\texttt{response\_ratio} $\leftarrow$ $N_{\mathrm{resp}} / N_{\mathrm{req}}$\;
              \If{$N_{\mathrm{resp}} = 0$}{
                  stats.\texttt{status} $\leftarrow$ \texttt{missing\_response}\;
              }
              \Else{
                  \If{resp.\texttt{response\_parse\_errors} $>$ 0}{
                      stats.\texttt{status} $\leftarrow$ \texttt{response\_parse\_errors}\;
                  }
                  \Else{
                      \If{resp.\texttt{decoder\_name\_mismatch\_count} $>$ 0}{
                          stats.\texttt{status} $\leftarrow$ \texttt{decoder\_name\_mismatch}\;
                      }
                      \Else{
                          stats.\texttt{status} $\leftarrow$ \texttt{ok}\;
                      }
                  }
              }
              AppendMatrixRow$(d, k, stats)$\;
          }
      }
      EmitReplayMatrix$(\mathbf{R})$\;
    \end{algorithm}
    Algorithm~\ref{alg:hw_decoder_pipeline} defines the core replay procedure used to evaluate decoder behavior under consistent hardware-derived inputs.

    \section{Xanadu Integration Case Study}
    \label{sec:xanadu_case}
    We use Xanadu public datasets as a first integration case.

    \subsection{Dataset Sources}
\begin{figure*}[t]
        \centering
        \includegraphics[width=1.85\columnwidth]{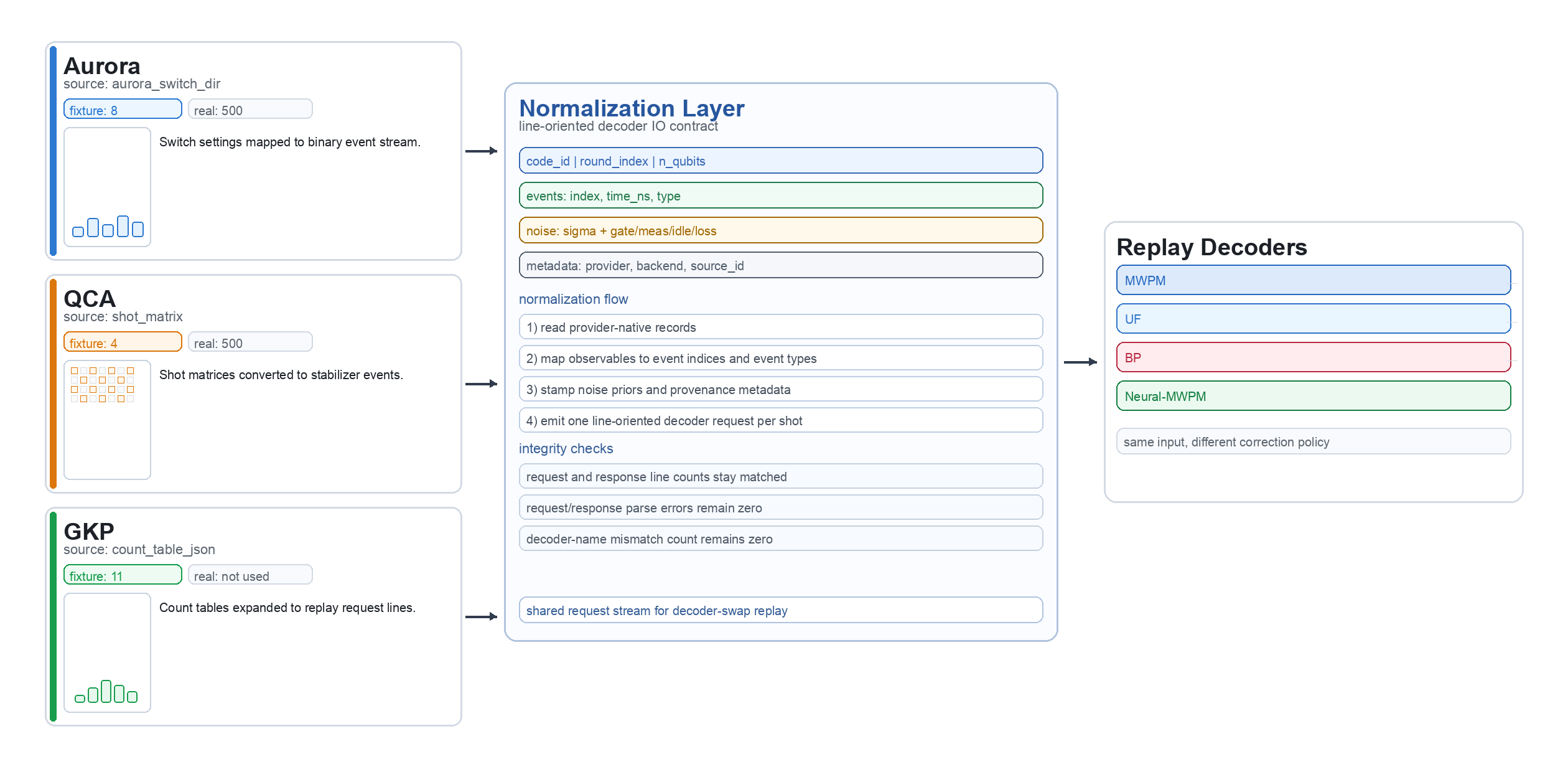}
        \caption{Design view of the three Xanadu data families used in this study. Aurora switch-setting traces, QCA shot-matrix samples, and GKP count-table summaries are normalized into one line-oriented decoder IO contract and replayed across MWPM, UF, BP, and neural-MWPM. Card annotations show request volumes used in fixture and real-slice stages when available.}
        \label{fig:xanadu_data_sources_design}
    \end{figure*}

    The fixture path covers four input families with distinct source formats:
    job-style records, switch-readout directories, shot matrices, and
    count summaries.
    For publicly released data, we cite both the underlying experimental studies
    and their dataset repositories: Aurora modular-photonic datasets
    \cite{aghaeerad2025aurora,xanadu2025aurora_data_repo}, QCA photonic-processor
    datasets \cite{madsen2022qca,xanadu2025qca_data_repo}, and GKP photonic-source
    datasets \cite{larsen2025integratedgkp,xanadu2025gkp_data_repo}. Real-data
    slices are drawn from these cited public sources.
    A compact design rendering of the three public data families used in this
    study (Aurora, QCA, GKP) and their normalization path to the shared decoder
    IO contract is shown in Fig.~\ref{fig:xanadu_data_sources_design}.
    Fixture request-generation coverage is summarized in Tab.~\ref{tab:request_manifest_fixture}.

    \begin{table}[t]
        \centering
        \caption{Fixture request-manifest summary generated by run `01\_prepare\_fixture\_requests.sh`.}
        \label{tab:request_manifest_fixture}
        \begin{tabular}{llr}
            \toprule
            Dataset & Request file & Request lines \\
            \midrule
            job & decoder\_requests.ndjson &        4 \\
            aurora & decoder\_requests\_aurora.ndjson &        8 \\
            gkp & decoder\_requests\_gkp.ndjson &       11 \\
            qca & decoder\_requests\_qca.ndjson &        4 \\
            \bottomrule
        \end{tabular}
    \end{table}

    \subsection{Mapping and Normalization}
    Mapping files translate provider observables to stabilizer-style event
    indices and event types used by decoders. During conversion, each shot is
    normalized to a single request line with:
    \begin{itemize}
        \item deterministic shot index and round index assignment,
        \item mapped syndrome events (\texttt{events}) with shared temporal fields,
        \item harmonized noise priors (\texttt{sigma}, gate, measurement, idle),
        \item metadata stamps for source format, backend, and dataset identifiers.
    \end{itemize}
    For Aurora fixtures we use binary-event mapping (\texttt{--aurora-binarize})
    so decoder inputs remain comparable to discrete-event paths.

    \subsection{Real-Data Slice Protocol}
    Real-data analysis uses controlled slice extraction so public datasets remain
    computationally tractable while preserving explicit source provenance and
    deterministic replay settings.

    \section{Methodology}
    \label{sec:setup}

    \subsection{Software and Versioning}
    Experiments in this study use LiDMaS+ \texttt{v1.2.0}.
    The replay engine is the compiled LiDMaS+ binary
    (\texttt{--decoder\_io\_replay}); orchestration and analysis are implemented
    as shell/Python scripts. With fixed inputs, reruns produce identical analysis outputs.

    \subsection{Actual Real Data Used}
    Tab.~\ref{tab:real_data_used_methods} lists the exact public-data slices
    used in this study's real-data replay matrix.

    \begin{table*}[t]
        \centering
        \caption{Actual real-data slices used for the optional hardware replay matrix in this study.}
        \label{tab:real_data_used_methods}
        \begin{tabular}{llcc}
            \toprule
            Dataset label & Source tag & Converted request file & Shots used \\
            \midrule
            aurora\_batch0\_qpu5 & Aurora decoder-demo (\texttt{batch\_0}, \texttt{qpu\_0--4}) & decoder\_requests\_aurora\_batch0\_qpu5.ndjson & 500 \\
            qca\_fig3b & QCA photonic processor (\texttt{fig3b}) & decoder\_requests\_qca\_fig3b.ndjson & 500 \\
            \bottomrule
        \end{tabular}
    \end{table*}
    Aurora source citations:
    \cite{aghaeerad2025aurora,xanadu2025aurora_data_repo}.
    QCA source citations:
    \cite{madsen2022qca,xanadu2025qca_data_repo}.

    \subsection{Decoder Set and Modes}
\begin{figure*}[t]
        \centering
        \includegraphics[width=1.85\columnwidth]{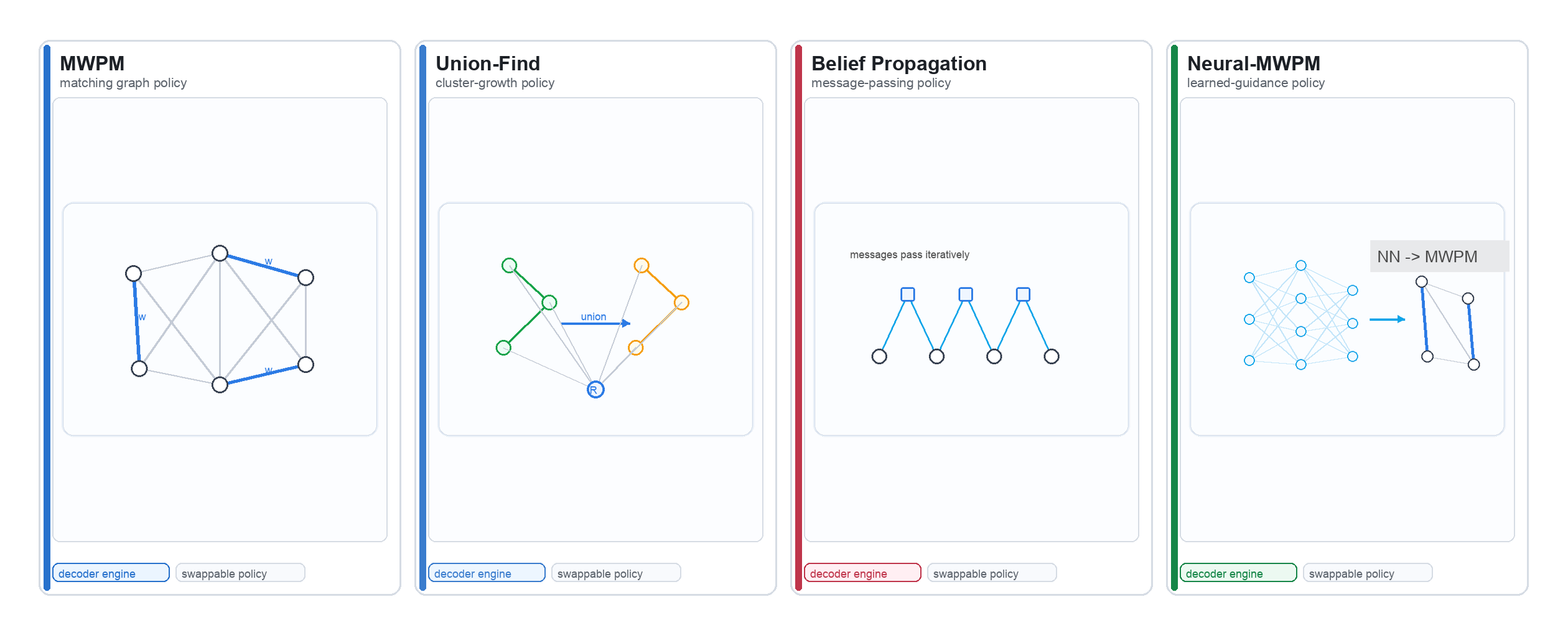}
        \caption{Merged standalone decoder designs used in replay: MWPM, Union-Find, Belief Propagation, and neural-guided MWPM. This panel provides a compact visual reference for the four engine options evaluated under matched request streams.}
        \label{fig:decoder_standalone_merged}
    \end{figure*}

    Default decoders are \texttt{mwpm}, \texttt{uf}, \texttt{bp}, and
    \texttt{neural\_mwpm} (when a model file is present). Per-decoder replay
    configurations keep the adapter contract fixed while mutating only
    decoder-specific fields. Decoder subsets can be selected at runtime,
    enabling controlled engine swaps under identical input traffic.
    A standalone visual summary of the four replay decoders is shown in
    Fig.~\ref{fig:decoder_standalone_merged}.

    \subsection{Metrics and Statistical Protocol}
    For each dataset \(d\) and decoder \(k\), let \(N^{\mathrm{req}}_{d,k}\) be
    request-line count and \(N^{\mathrm{resp}}_{d,k}\) be response-line count.
    We report:
    \begin{align}
        \rho_{d,k} &= \frac{N^{\mathrm{resp}}_{d,k}}{N^{\mathrm{req}}_{d,k}}
        && \text{(response ratio)} \\
        w_{d,k} &= \frac{N^{\mathrm{warn}}_{d,k}}{N^{\mathrm{resp}}_{d,k}}
        && \text{(no-syndrome warning rate)} \\
        f_{d,k} &= \frac{1}{N^{\mathrm{resp}}_{d,k}} \sum_{i=1}^{N^{\mathrm{resp}}_{d,k}}
        \left|\mathrm{flips}_{i}\right|
        && \text{(average flip count)} \\
        \eta_{d,k} &= \frac{N^{\mathrm{nonempty\_flip}}_{d,k}}{N^{\mathrm{resp}}_{d,k}}
        && \text{(nonempty flip rate)}.
    \end{align}

    We additionally track request/response parse-error counts, decoder-name
    mismatch counts, average \(S_X/S_Z\) syndrome diagnostics, and unique flipped
    qubit support. Aggregates in Results are request-count weighted unless noted
    otherwise. The statistical emphasis is cross-decoder behavioral separation
    under matched inputs; confidence-interval and threshold sweeps are deferred
    to larger-window studies.

    \subsection{Control and Ablation Protocol}
    To separate decoder-policy effects from provider-specific effects, we add a
    matched-sparsity control protocol alongside real-data replay. First, we
    compute residual-syndrome quality metrics for fixture and real slices:
    syndrome-satisfaction rate and residual nonzero rate after applying decoder
    corrections to the observed syndrome stream. Second, we generate non-photonic
    synthetic heldout requests whose syndrome sparsity is approximately matched to
    the real-slice distributions, replay the same decoder set, and evaluate both
    residual-syndrome metrics and logical-failure rates using ground-truth labels
    embedded in request metadata. This design does not replace real-data
    evaluation; it provides a control axis for testing whether observed decoder
    rankings are tied to photonic-provider structure or persist under
    sparsity-matched synthetic inputs.

    \section{Results}
    \label{sec:results}
    We evaluate whether a single hardware-to-decoder interface can preserve input
    semantics across heterogeneous sources while enabling valid decoder
    comparison. In this Xanadu-centered study, fixture and real slices both
    achieved complete request-to-response coverage with zero parser failures, and
    decoder correction signatures remained clearly separated under matched
    requests. The results below report interface correctness, decoder behavior,
    residual-quality tradeoffs, and robustness across sparsity regimes.

    \subsection{Conversion Correctness and Schema Compliance}
\begin{figure*}[t]
        \centering
        \includegraphics[width=1.85\columnwidth]{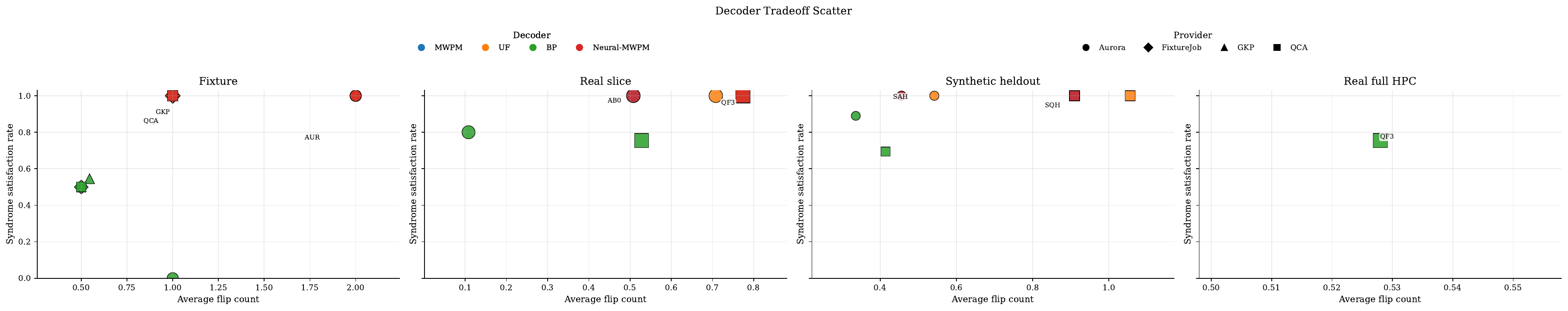}
        \caption{Decoder tradeoff scatter across fixture, real-slice, and synthetic-heldout scopes. Point color denotes decoder, marker denotes provider/dataset, and point size scales with request volume/nonempty-flip activity. Short labels are dataset codes: J (job), AUR (aurora), GKP (gkp), QCA (qca), AB0 (aurora\_batch0\_qpu5), QF3 (qca\_fig3b), SAH (synth\_aurora\_batch0\_qpu5\_heldout), and SQH (synth\_qca\_fig3b\_heldout). This visual highlights that lower correction volume does not necessarily imply better correction quality.}
        \label{fig:decoder_tradeoff_scatter}
    \end{figure*}

\begin{figure*}[t]
        \centering
        \includegraphics[width=1.85\columnwidth]{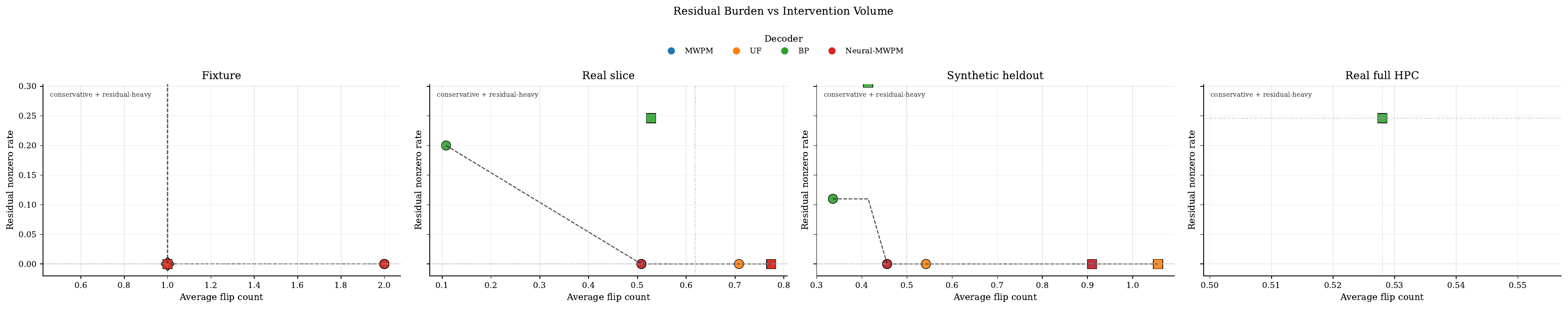}
        \caption{Residual burden versus intervention volume by scope. The dashed envelope gives a Pareto-style guide for conservative-vs-effective correction behavior. BP appears in the low-intervention region but with higher residual burden than MWPM-like decoders in the tested windows.}
        \label{fig:residual_vs_intervention}
    \end{figure*}

    Decoder replay quality on fixture datasets is summarized in
    Tab.~\ref{tab:decoder_matrix_fixture}.
    Visual summaries for decoder separability, intervention-versus-residual
    tradeoff, and engine-swap invariance are shown in
    Figs.~\ref{fig:decoder_tradeoff_scatter},
    \ref{fig:residual_vs_intervention}, and
    \ref{fig:engine_swap_consistency}.

\begin{figure*}[t]
        \centering
        \subfloat[Average flip count vs sparsity bucket]{
            \includegraphics[width=0.48\textwidth]{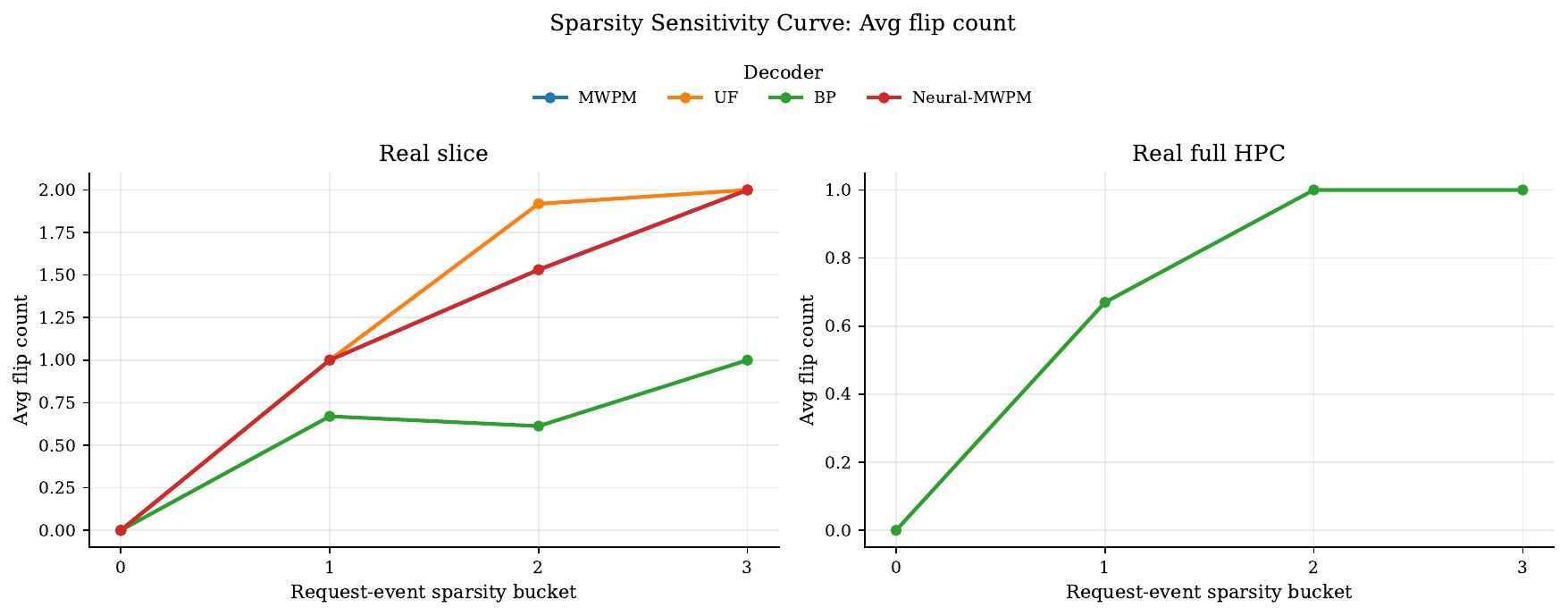}
        }
        \hfill
        \subfloat[Nonempty flip rate vs sparsity bucket]{
            \includegraphics[width=0.48\textwidth]{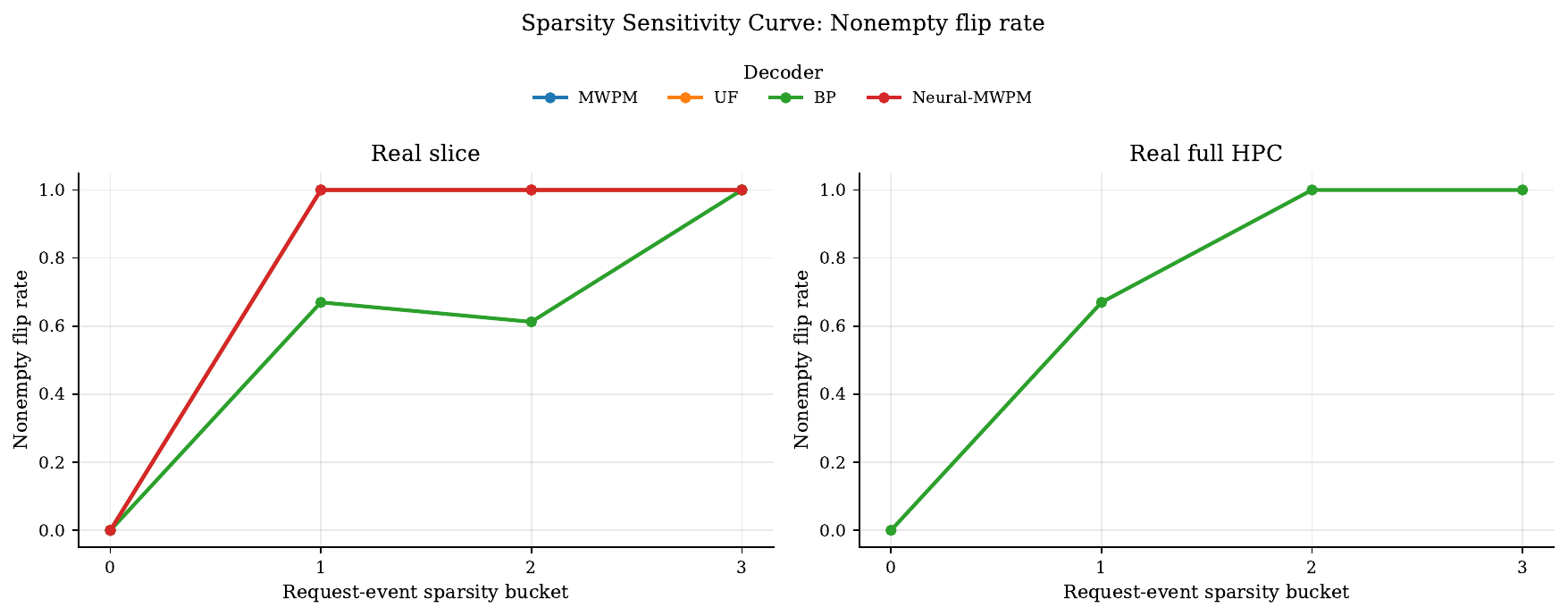}
        }
        \caption{Sparsity-sensitivity curves with requests bucketed by event-count sparsity. Decoder separation varies systematically with sparsity regime, supporting the interpretation that differences are amplified in sparse-input windows.}
        \label{fig:sparsity_sensitivity_curves}
    \end{figure*}

\begin{figure*}[t]
        \centering
        \includegraphics[width=1.85\columnwidth]{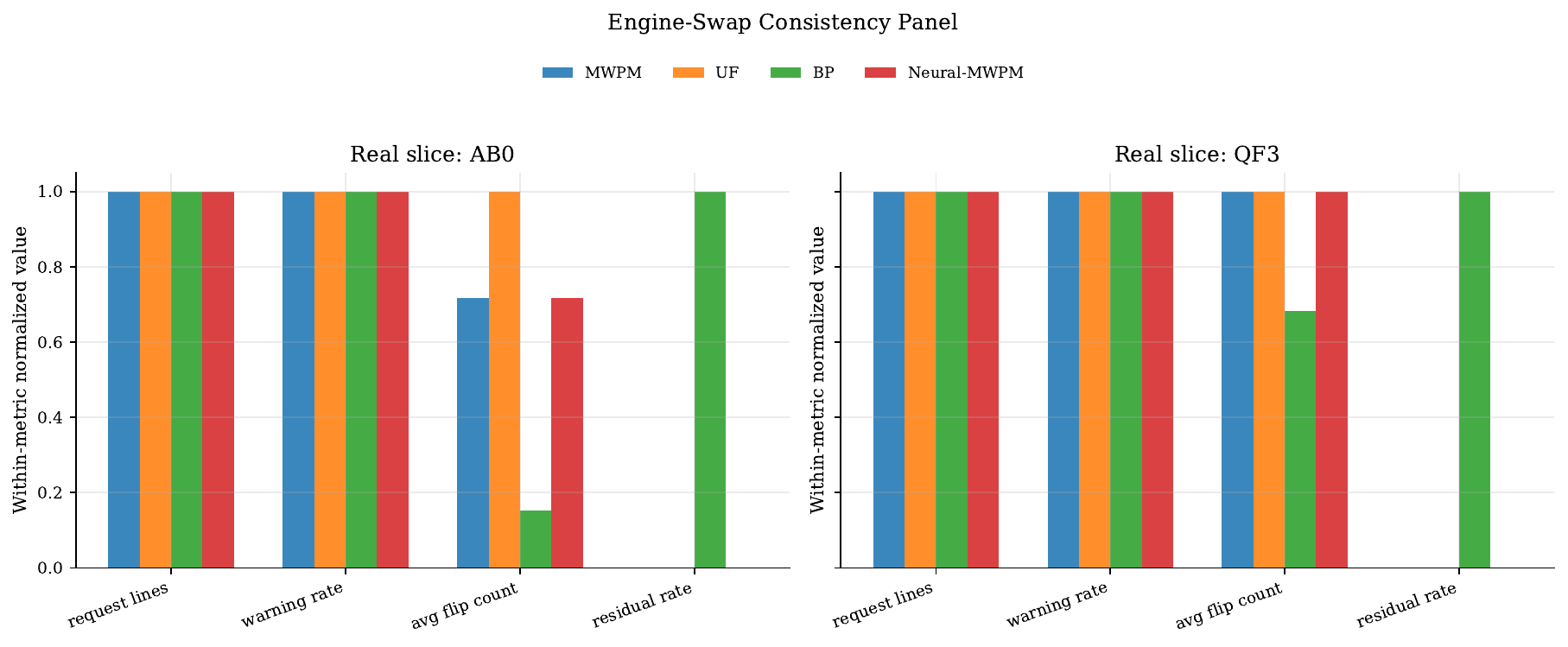}
        \caption{Engine-swap consistency panel on matched request streams. Titles use short dataset codes to reduce clutter: AB0 denotes aurora\_batch0\_qpu5 and QF3 denotes qca\_fig3b. Request counts and warning rates remain invariant across decoder selection, while correction volume and residual metrics differ by decoder policy. This supports the ``same input contract, swappable decoder engine'' claim.}
        \label{fig:engine_swap_consistency}
    \end{figure*}

\begin{figure*}[t]
        \centering
        \includegraphics[width=2\columnwidth]{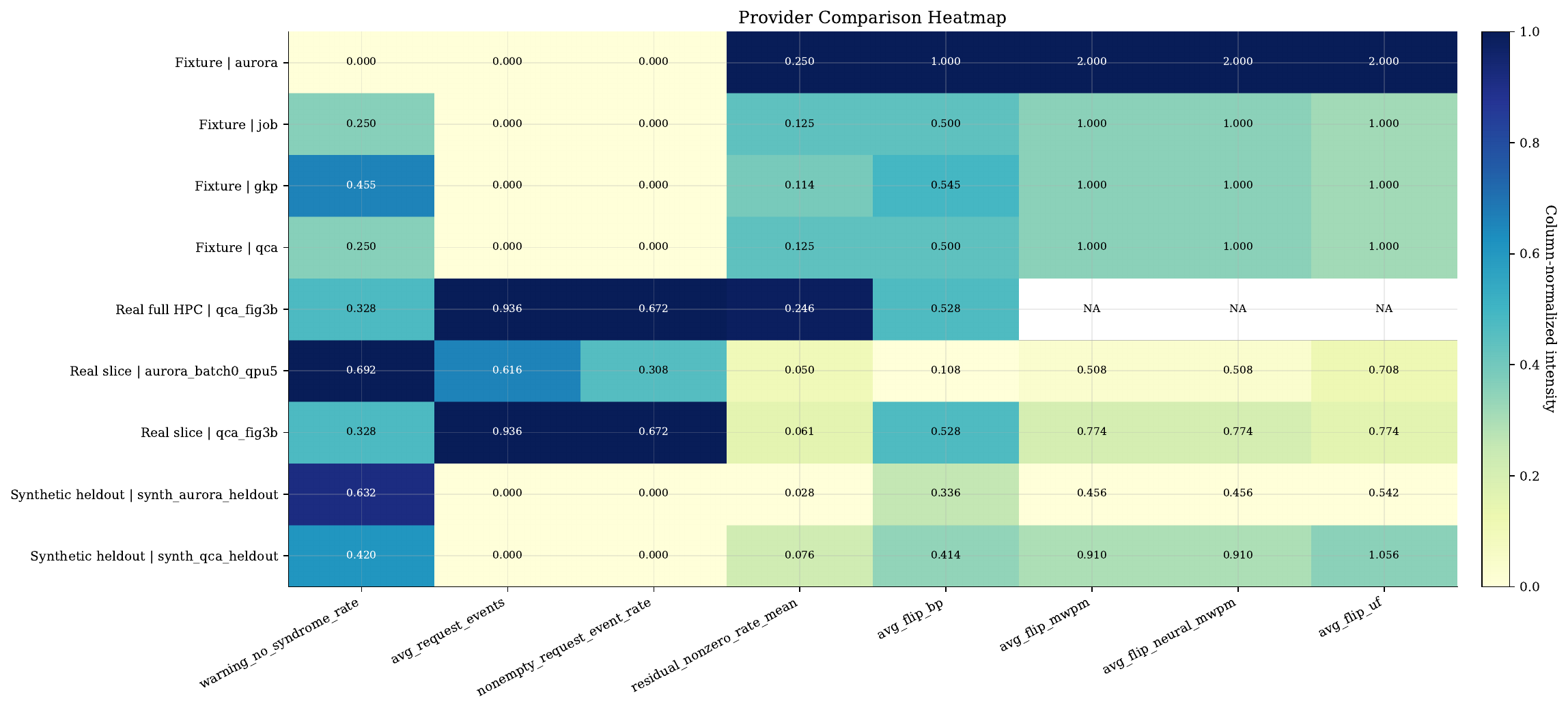}
        \caption{Provider comparison heatmap under one visual grammar. Rows are dataset/provider slices; columns combine warning and sparsity structure with decoder-specific intervention burden. This layout is extensible to additional providers (e.g., Quandela, Google) without changing analysis semantics.}
        \label{fig:provider_comparison_heatmap}
    \end{figure*}

    Figs.~\ref{fig:decoder_tradeoff_scatter}--\ref{fig:provider_comparison_heatmap}
    form one linked diagnostic set. Fig.~\ref{fig:decoder_tradeoff_scatter}
    shows that lower intervention volume does not guarantee higher
    syndrome-satisfaction quality: BP is typically low-intervention, while
    MWPM-like decoders are often higher on satisfaction. Fig.~\ref{fig:residual_vs_intervention}
    makes the same point on residual burden, highlighting the intervention-versus-residual
    tradeoff directly. Fig.~\ref{fig:sparsity_sensitivity_curves} shows that this
    separation is regime dependent and changes with sparsity bucket. Fig.~\ref{fig:engine_swap_consistency}
    confirms engine-swap invariance: request counts and warning rates stay fixed
    across decoders, while correction metrics change. Fig.~\ref{fig:provider_comparison_heatmap}
    places provider structure and decoder behavior under one normalized grammar
    that can extend to new providers without changing metric semantics.

    Correctness was evaluated at syntactic and semantic levels. Syntactically,
    requests and responses parsed cleanly with line-level response ratio exactly
    one. Semantically, decoder identifiers matched replay targets and error
    counters stayed zero while sparsity structure was preserved. Fixture replay
    produced 108/108 valid responses; real slices produced 4000/4000 valid
    responses. In every matrix cell, \texttt{response\_ratio}=1.0,
    \texttt{request\_parse\_errors}=0, \texttt{response\_parse\_errors}=0, and
    \texttt{error\_count}=0.

    The no-syndrome warning signal is informative here because it measures input
    sparsity at replay time, not parser failure. In the fixture case, warning
    rates are dataset-dependent and decoder-invariant: \texttt{aurora}=0.000,
    \texttt{job}=0.250, \texttt{qca}=0.250, and \texttt{gkp}=0.455. Weighted by
    request count, this yields an overall fixture warning rate of
    \(7/27 \approx 0.259\). In the real slice case, the same metric separates
    two very different data regimes: \texttt{aurora\_batch0\_qpu5}=0.692 and
    \texttt{qca\_fig3b}=0.328, for a weighted mean of \(510/1000=0.510\). The
    key point is that this variation follows dataset structure, not decoder
    choice, confirming that input characteristics survive conversion and are not
    overwritten by replay settings.

    Event-density metrics support the same interpretation. In fixtures, Aurora is
    the densest stream (\texttt{avg\_request\_events}=3.0, nonempty rate \(=1.0\)),
    while GKP is much sparser (\texttt{avg\_request\_events}=1.091, nonempty rate
    \(=0.545\)). In real slices, Aurora remains sparser
    (\texttt{avg\_request\_events}=0.616, nonempty rate \(=0.308\)) than QCA
    (\texttt{avg\_request\_events}=0.936, nonempty rate \(=0.672\)).

    Finally, there were no decoder-name mismatches in any replay cell
    (\texttt{decoder\_name\_mismatch\_count}=0 throughout). This matters for
    large benchmark grids where silent decoder-routing mistakes can invalidate
    comparisons. The combination of full line coverage, zero parsing failures,
    zero decoder mismatches, and stable dataset-dependent warning structure
    indicates that the converter+replay contract is functioning as intended for
    both fixture and sampled real data. In practical terms, the transport and
    normalization layer is stable enough to support decoder-level interpretation.

    \subsection{Decoder Performance Across Hardware-Derived Inputs}
    We next analyze decoder behavior under identical converted inputs. The
    principal comparison metric is
    \texttt{avg\_flip\_count} (average number of proposed correction flips per
    request), together with \texttt{nonempty\_flip\_rate} and
    \texttt{unique\_flip\_qubits}. These are operational diagnostics rather than
    final logical-failure metrics: they quantify correction ``style'' under the
    same syndrome stream. A lower flip count can indicate conservative behavior
    that may reduce over-correction, but by itself does not prove better logical
    accuracy. We therefore interpret these values as signatures of decoder policy
    under matched conditions.

    On fixture data, MWPM, UF, and neural-MWPM are nearly indistinguishable in
    this run, while BP is consistently less aggressive. At the dataset level:
    for \texttt{job}, MWPM/UF/neural each report \texttt{avg\_flip\_count}=1.0,
    BP reports 0.5; for \texttt{aurora}, MWPM/UF/neural each report 2.0, BP
    reports 1.0; for \texttt{gkp}, MWPM/UF/neural each report 1.0, BP reports
    0.545; and for \texttt{qca}, MWPM/UF/neural each report 1.0, BP reports
    0.5. This corresponds to BP reductions of 50\% (job), 50\% (aurora),
    45.5\% (gkp), and 50\% (qca) relative to MWPM.

    Aggregating across all fixture requests with request-count weighting
    highlights the same pattern. MWPM, UF, and neural-MWPM each yield a weighted
    \texttt{avg\_flip\_count} of \(1.296\), while BP yields \(0.667\). Thus BP
    uses roughly 48.6\% fewer flips overall in the fixture matrix. The
    nonempty-flip profile is similar: MWPM/UF/neural average
    \texttt{nonempty\_flip\_rate}\(\approx 0.741\), BP averages
    \(\approx 0.667\). This means BP reduces both absolute correction volume and
    the fraction of requests that trigger nontrivial corrections.

    The qubit-support metric (\texttt{unique\_flip\_qubits}) shows the same
    directional behavior. In the fixture Aurora row, BP touches one unique qubit
    while MWPM/UF/neural each touch four; in fixture job and qca, BP touches two
    while MWPM/UF/neural each touch three; in fixture gkp, BP and UF touch two,
    MWPM touches two, and neural touches three. The data therefore suggest that
    BP's reduced flip volume is not only a magnitude change but often a support
    contraction as well.

    Real-data slices preserve the same qualitative ranking while expanding the
    dynamic range. For \texttt{aurora\_batch0\_qpu5} (500 requests), MWPM and
    neural-MWPM each report \texttt{avg\_flip\_count}=0.508, UF reports 0.708,
    and BP reports 0.108. Compared with MWPM, BP reduces correction volume by
    \(0.400\) flips per request, i.e., 78.7\%. For \texttt{qca\_fig3b}
    (500 requests), MWPM/UF/neural each report 0.774 while BP reports 0.528,
    corresponding to a 31.8\% reduction versus MWPM. Thus BP remains the most
    conservative decoder in both real-slice families, but the size of the gap is
    dataset-dependent.

    Weighted across both real slices, MWPM and neural-MWPM each produce
    \texttt{avg\_flip\_count}=0.641, UF produces 0.741, and BP produces 0.318.
    Relative to MWPM, BP cuts weighted correction volume by 50.4\%; relative to
    UF, BP cuts it by 57.1\%. This ranking is mirrored by
    \texttt{nonempty\_flip\_rate}: MWPM/UF/neural each average 0.49, BP averages
    0.318. In other words, under real sampled data BP both intervenes less often
    and flips fewer qubits when it does intervene.

    Figs.~\ref{fig:decoder_tradeoff_scatter},
    \ref{fig:residual_vs_intervention}, and
    \ref{fig:sparsity_sensitivity_curves} make two additional points clear.
    First, warning-rate behavior is decoder-flat because warnings are properties
    of incoming syndrome sparsity and request structure. Second, correction
    behavior is decoder-structured, and that structure is consistent across
    fixture, real, and synthetic scopes. Fig.~\ref{fig:provider_comparison_heatmap}
    provides a compact provider-level view of this separation pattern under one
    visual grammar. The combination of
    decoder-invariant input diagnostics and decoder-variant correction diagnostics
    is exactly what a useful benchmark interface should produce.

    One practical consequence is that decoders can be swapped ``as an engine''
    without retooling ingestion. In this run, the same normalized requests were
    replayed through MWPM, UF, BP, and neural-MWPM by configuration only.
    Because the IO and replay contract stayed fixed while decoder signatures
    changed in stable ways, the workflow supports A/B decoder studies and
    deployment-style decoder selection from a shared hardware-derived request
    stream. That is the core operational advantage targeted by this study.

    \subsection{Residual-Syndrome and Control-Ablation Results}
    We next test whether low flip volume corresponds to stronger correction
    quality. On real slices, MWPM, UF, and neural-MWPM each achieved
    syndrome-satisfaction rate \(=1.000\), while BP achieved a weighted
    syndrome-satisfaction rate of \(0.777\) (Aurora \(0.800\), QCA \(0.754\)),
    with weighted residual nonzero rate \(=0.223\). This shows that BP's lower
    correction volume in the main matrix is accompanied by a higher residual
    syndrome burden in this setting.

    To test whether this pattern is purely photonic-provider specific, we ran a
    non-photonic synthetic heldout control with approximately matched syndrome
    sparsity. The separation persisted: MWPM/UF/neural-MWPM remained at
    syndrome-satisfaction rate \(=1.000\), while BP was \(0.792\) (weighted).
    However, heldout logical-failure rates were close across decoders in this
    control window: BP \(0.042\), MWPM \(0.045\), UF \(0.045\), and neural-MWPM
    \(0.046\). The practical interpretation is that ``best'' depends on the
    reported metric: BP is most conservative in intervention volume, but not
    strongest on residual-syndrome clearance in these runs. This control-study
    axis directly improves causal interpretability of decoder comparisons.

    \subsection{Throughput, Scaling, and Reproducibility}
    Optional real-data slice replay metrics (Aurora/QCA subsets) are summarized in
    Tab.~\ref{tab:decoder_matrix_real}.

    Throughput in this study is measured first as completion integrity: whether
    each request line produces exactly one valid response line in every matrix
    cell. On this criterion, completion reached 100\% in all runs.
    The fixture matrix completed 108/108 lines and the real-slice matrix
    completed 4000/4000 lines, each with \texttt{response\_ratio}=1.0 and zero
    parse failures. Incomplete replay can bias decoder comparisons by silently
    dropping difficult cases; this behavior was not observed here.

    Re-running the analysis produced identical SHA-256 fingerprints for fixture
    and real-slice artifacts; no reported values changed between runs. Full
    fingerprints are listed in Appendix~\ref{app:repro_hashes} and
    Tab.~\ref{tab:sha256_fingerprints}.

    In terms of scaling interpretation, this study's results are based on a
    24-cell replay matrix (16 fixture cells + 8 real-slice cells) and
    intentionally prioritize interface correctness over large-\(N\) timing
    sweeps. Even so, the run structure already exposes several properties that
    matter for scale-up: composable dataset slicing, decoder-parallel matrix
    expansion, and deterministic artifact regeneration. These properties make it
    straightforward to expand shot windows and provider coverage while keeping
    the same evaluation contract.

    \begin{table*}[t]
        \centering
        \caption{Decoder replay matrix summary for fixture datasets (`03\_analyze\_decoder\_matrix.sh`).}
        \label{tab:decoder_matrix_fixture}
        \begin{tabular}{llcccc}
            \toprule
            Dataset & Decoder & Request lines & Response lines & No-syndrome rate & Avg flip count \\
            \midrule
            job & mwpm & 4 & 4 & 0.250 & 1.000 \\
            job & uf & 4 & 4 & 0.250 & 1.000 \\
            job & bp & 4 & 4 & 0.250 & 0.500 \\
            job & neural\_mwpm & 4 & 4 & 0.250 & 1.000 \\
            aurora & mwpm & 8 & 8 & 0.000 & 2.000 \\
            aurora & uf & 8 & 8 & 0.000 & 2.000 \\
            aurora & bp & 8 & 8 & 0.000 & 1.000 \\
            aurora & neural\_mwpm & 8 & 8 & 0.000 & 2.000 \\
            gkp & mwpm & 11 & 11 & 0.455 & 1.000 \\
            gkp & uf & 11 & 11 & 0.455 & 1.000 \\
            gkp & bp & 11 & 11 & 0.455 & 0.545 \\
            gkp & neural\_mwpm & 11 & 11 & 0.455 & 1.000 \\
            qca & mwpm & 4 & 4 & 0.250 & 1.000 \\
            qca & uf & 4 & 4 & 0.250 & 1.000 \\
            qca & bp & 4 & 4 & 0.250 & 0.500 \\
            qca & neural\_mwpm & 4 & 4 & 0.250 & 1.000 \\
            \bottomrule
        \end{tabular}
    \end{table*}

    \begin{table*}[t]
        \centering
        \caption{Decoder replay matrix summary for optional real-data slices (`05\_analyze\_real\_data\_slice.sh`).}
        \label{tab:decoder_matrix_real}
        \begin{tabular}{llcccc}
            \toprule
            Dataset & Decoder & Request lines & Response lines & No-syndrome rate & Avg flip count \\
            \midrule
            aurora\_batch0\_qpu5 & mwpm & 500 & 500 & 0.692 & 0.508 \\
            aurora\_batch0\_qpu5 & uf & 500 & 500 & 0.692 & 0.708 \\
            aurora\_batch0\_qpu5 & bp & 500 & 500 & 0.692 & 0.108 \\
            aurora\_batch0\_qpu5 & neural\_mwpm & 500 & 500 & 0.692 & 0.508 \\
            qca\_fig3b & mwpm & 500 & 500 & 0.328 & 0.774 \\
            qca\_fig3b & uf & 500 & 500 & 0.328 & 0.774 \\
            qca\_fig3b & bp & 500 & 500 & 0.328 & 0.528 \\
            qca\_fig3b & neural\_mwpm & 500 & 500 & 0.328 & 0.774 \\
            \bottomrule
        \end{tabular}
    \end{table*}

    \subsection{Sensitivity and Robustness Checks}
    We performed robustness checks to test whether observed decoder differences
    persist under shifts in syndrome sparsity and dataset family. The first check is
    ``input-sensitivity separation'': warning rates should move with data source
    rather than decoder. This condition holds across both fixture and real
    slices. For every dataset, warning rates are identical across decoders, while
    average flip counts differ by decoder. Therefore decoder rankings are not an
    artifact of a decoder-specific input preprocessing path.

    The second check is ``cross-regime rank stability.'' We compare two regimes:
    sparse-event slices (e.g., \texttt{aurora\_batch0\_qpu5}, nonempty request
    event rate \(=0.308\)) and denser slices (e.g., \texttt{qca\_fig3b}, nonempty
    request event rate \(=0.672\); fixture aurora, nonempty rate \(=1.0\)). BP
    remains the lowest-flip decoder in every regime, while MWPM and neural-MWPM
    remain closely aligned. UF tracks MWPM on fixture data but shifts upward in
    real aurora (0.708 vs 0.508), showing that regime changes can alter
    near-ties even when global ordering is mostly stable.

    The third check is ``magnitude stability under dataset shifts.'' BP's
    reduction relative to MWPM is largest on sparse real aurora (78.7\%), smaller
    on real qca (31.8\%), and intermediate on fixtures (45.5--50\%). This range
    is expected: decoder policy differences are amplified when input streams are
    sparse and potentially damped when event density rises. The important point
    for this study is that directionality does not invert across datasets.

    We also examined failure-mode robustness. Across all matrix cells there were
    no parser failures, no explicit decoder errors, and no decoder-name mismatch
    events. This does not prove absence of semantic edge cases, but it does
    establish a clean baseline for larger stress tests. In practical use, this
    means researchers can expand shot windows or add provider datasets without
    first resolving schema-instability issues.

    Remaining robustness work is straightforward: shot-window sweeps beyond 500
    samples per real subset, bootstrap confidence intervals for correction
    metrics, and mapping-ablation experiments that perturb normalization choices.
    The present results should be read as a stable integration baseline with
    informative decoder signatures, not yet as a final statistical claim about
    logical error suppression on hardware.

    \section{Discussion}
    \label{sec:discussion}
    The primary result of this study is an architecture-level separation between
    hardware-origin input structure and decoder policy response. With one
    contract and matched replay controls, input diagnostics remain
    decoder-invariant while correction behavior remains decoder-dependent. This
    is the condition required for meaningful cross-decoder interpretation in
    hardware-facing QEC studies.

    The central scientific insight is regime dependence. Decoder ranking is not
    global; it shifts with syndrome sparsity and event structure. In our runs,
    BP consistently minimizes intervention volume, while MWPM-family decoders
    apply larger corrections and generally clear more syndrome. This is the
    dominant interpretability axis for the study.

    The intervention-versus-residual tradeoff is therefore explicit: lower flip
    count can come with higher residual burden, and higher correction volume can
    reduce residual syndrome at additional intervention cost. For architecture
    decisions, this tradeoff is more informative than any single scalar metric.
    It should be treated as a policy surface, not as a one-dimensional ranking.

    The same structure appears when comparing MWPM, UF, and neural-MWPM.
    MWPM and neural-MWPM remain close in several settings, while UF is close in
    some regimes and separates in others. This pattern supports a practical
    design rule: decoder choice should be conditioned on expected operating
    regime (especially sparsity profile), latency envelope, and correction-cost
    budget.

    The present benchmarks are derived from stabilizer-style discrete-variable frameworks; however, the observed decoder-policy tradeoffs extend directly to GKP-based CV-DV systems, where discretized syndrome extraction produces structurally analogous decoding behavior \cite{gkp2001,tzitrin2021static,larsen2025integratedgkp}. As these platforms scale, decoder benchmarking must remain faithful to hardware semantics while preserving strict comparability across decoder engines. The LiDMaS+ design directly supports this requirement.

    From an operational perspective, the architecture enables engine-style
    decoder swapping: request streams are fixed, decoder configuration changes,
    and measured differences can be attributed to policy rather than ingestion
    drift. This supports a controlled A/B loop for hardware-in-the-loop studies
    and lowers the cost of testing new decoder backends.

    The matched-sparsity control strengthens causal interpretation of decoder
    behavior. When ranking patterns persist across real and synthetic matched
    conditions, confidence in policy-level effects increases. When they diverge,
    the divergence identifies interactions between decoder behavior and
    data-generation context. This is the correct level of interpretability for
    architecture selection in heterogeneous hardware datasets.

    The architecture implication is direct: fixed-decoder deployment is often
    suboptimal across operating regimes. Regime-aware adaptive or hybrid decoder
    strategies are better aligned with observed behavior, for example using
    conservative policies in low-risk sparse windows and more aggressive
    correction policies when residual control dominates system objectives.
    LiDMaS+ provides the execution and measurement substrate needed to design and
    validate such switching policies under consistent hardware-derived inputs.

    Finally, deterministic regeneration and parser-level correctness remain
    essential because they protect the scientific meaning of decoder comparisons.
    With those guarantees in place, expanding shot windows, provider coverage,
    and uncertainty quantification becomes a scientific scaling problem rather
    than an integration-risk problem.

    \begin{table*}[t]
        \centering
        \caption{SHA-256 fingerprints used for deterministic-rerun verification.}
        \label{tab:sha256_fingerprints}
        \begin{tabular}{ll}
            \toprule
            Output & SHA-256 fingerprint \\
            \midrule
            Fixture replay summary & \texttt{ab429199121e9f45 c51f0d6739fff00e} \\
            & \texttt{498436f524fb0a47 217ddbb66c2ffd87} \\
            Real-slice replay summary & \texttt{581fd2368cd64b32 36ba7d244ccabc0e} \\
            & \texttt{9be59392448aec37 dfcc052384d30582} \\
            \bottomrule
        \end{tabular}
    \end{table*}

    \subsection{Limitations}
    Several limitations define the current maturity boundary.

    First, provider coverage is intentionally narrow. The implementation has been
    refactored for a multi-provider future, but the present end-to-end case study
    is Xanadu-only. This means schema decisions and mapping assumptions are
    stress-tested on one ecosystem first, not yet on a fully diverse set of
    hardware telemetry formats.

    Second, real-data evaluation is slice-based. The study reports 500-shot
    subsets for two public datasets to keep run cost manageable during pipeline
    development. These slices are adequate for integration validation and
    decoder-signature discovery, but not sufficient for final statistical claims
    about tail behavior or long-horizon stability.

    Third, reported metrics now include residual-syndrome quality diagnostics for
    fixture/real slices and heldout logical-failure rates for synthetic controls.
    However, we do not yet report full logical error rate curves on real hardware
    slices, confidence intervals across independent windows, or threshold-style
    scaling analyses. We therefore treat decoder differences as policy signatures
    under fixed inputs, not as definitive superiority claims.

    Fourth, throughput is currently represented by completion integrity and
    deterministic artifact regeneration, rather than wall-clock benchmarking under
    controlled hardware and thread settings. This is appropriate for an
    integration study, but a production deployment study should include explicit
    latency and throughput-service objectives.

    Fifth, neural-MWPM parity with MWPM in these runs indicates either equivalent
    behavior under the active model/configuration or insufficient data diversity
    to expose differences. A broader model-ablation set is required before making
    stronger claims about neural guidance in this hardware-derived regime.

    \section{Conclusion}
    \label{sec:conclusion}
    We presented a hardware-to-logical-to-decoder architecture in LiDMaS+ and
    demonstrated it on a full Xanadu integration path. Across fixture and
    sampled real data, replay coverage was complete (108/108 and 4000/4000
    lines), with zero parse failures and zero decoder-name mismatches.

    Under matched inputs, decoder behavior is clearly separable and
    regime-dependent. Weighted fixture \texttt{avg\_flip\_count} values were
    1.296 (MWPM), 1.296 (UF), 0.667 (BP), and 1.296 (neural-MWPM). Weighted
    real-slice values were 0.641 (MWPM), 0.741 (UF), 0.318 (BP), and 0.641
    (neural-MWPM), with weighted real-slice \texttt{nonempty\_flip\_rate}
    values of 0.490, 0.490, 0.318, and 0.490, respectively.

    These results establish the core interpretability statement of this work:
    decoder selection is a tradeoff between intervention cost and residual error.
    BP is intervention-conservative and often leaves higher residual burden;
    MWPM-family decoders intervene more aggressively and generally reduce
    residual syndrome more strongly. This tradeoff, rather than any single
    metric, should guide architecture-level decoder policy.

    For photonic GKP-oriented QEC programs, this has direct deployment
    implications. As hardware regimes shift in sparsity and noise structure,
    static decoder choice is unlikely to remain optimal. Regime-aware adaptive
    or hybrid decoder strategies become the natural next step, and LiDMaS+
    provides the measurement framework needed to design and validate them under
    controlled hardware-derived inputs.

    Future work should scale this architecture across additional providers,
    larger real-data windows, stronger uncertainty quantification, and explicit
    latency-throughput constraints. With these extensions, the same framework can
    support both rigorous comparative research and operational decoder policy
    selection for hybrid CV-DV fault-tolerant quantum computing.

    \section*{Author Contributions}

    D.D.K.W: conceptualization, methodology, validation and visualization, software, writing – original draft, review \& editing. C.O: methodology, validation and visualization, software, writing – original draft, review \& editing. L.G: methodology, validation and visualization, software, writing – original draft, review \& editing. S.G: methodology, validation and visualization, software, writing – original draft, review \& editing.

    \section*{Acknowledgment(s)}

    The authors acknowledge contributors, users, and reviewers who provided
    feedback on decoding workflows, reproducibility scripts, and
    documentation quality. Any opinions, findings, conclusions, or recommendations expressed in this research are those of the author(s) and do not necessarily reflect the views of their respective affiliations.

    \section*{Data \& Code Availability}
    The data generated and analyzed during the present study are included within this study. Supplementary codes developed from \texttt{LiDMaS+} simulator are provided as supplementary material and accessible on \href{https://github.com/DennisWayo/lidmas_cpp}{GitHub} to ensure transparency and reproducibility.

    \section*{Funding}
    This research was not funded.

    \section*{Disclosure statement}

    No potential conflict of interest was reported by the author(s).

    \appendix
    \section{Full Reproducibility Protocol}
    \label{app:repro_hashes}
    All commands below are run from repository root.

    \textbf{Install and baseline workflow}
    \begin{verbatim}
python -m pip install --upgrade lidmas
cmake -S . -B build
cmake --build build -j
./examples/paper_runs/paper_03/run_all.sh
    \end{verbatim}

    \textbf{Real-data slice workflow (same settings as this study)}
    \begin{verbatim}
export LIDMAS_RUN_REAL_DATA=1
export LIDMAS_HW_DATASETS=aurora_min,qca_fig3b
export LIDMAS_HW_MAX_SHOTS=500
./examples/paper_runs/paper_03/run_all.sh
    \end{verbatim}

    \textbf{Explicit artifact refresh}
    \begin{verbatim}
./examples/paper_runs/paper_03/06_sync_tables_to_tex.sh
./examples/paper_runs/paper_03/07_generate_figures.sh
    \end{verbatim}

    \textbf{Deterministic-rerun verification}
    \begin{verbatim}
RUN=./examples/paper_runs/paper_03
${RUN}/03_analyze_decoder_matrix.sh
${RUN}/05_analyze_real_data_slice.sh
OUT=${RUN}/results
DECODER_CSV=${OUT}/03_decoder_matrix_analysis/
table_decoder_matrix.csv
REAL_CSV=${OUT}/05_real_data_analysis/t
able_real_data_decoder_matrix.csv
shasum \
  -a 256 \
  "${DECODER_CSV}"
shasum \
  -a 256 \
  "${REAL_CSV}"
    \end{verbatim}

    \bibliographystyle{apsrev4-2}
    \bibliography{lidmas}

\end{document}

%% file: preamble.tex
\usepackage{nag}
\usepackage{graphicx}
\usepackage{amsthm}
\usepackage{tabulary}
\usepackage{qcircuit}
\usepackage{mathtools}
\usepackage{bm}
\usepackage{braket}
\usepackage{datetime}
\usepackage{stackrel}
\usepackage{dsfont}
\usepackage{qcircuit}
\usepackage[english]{babel}
\usepackage{units}

\usepackage{tikz}
\usetikzlibrary{backgrounds,decorations.pathreplacing}

\usepackage{chngcntr}
\counterwithout{equation}{section}

\usepackage[normalem]{ulem}
\usepackage{slashed}
\usepackage{soul}

\usepackage{xcolor}
\usepackage{amssymb}
\usepackage{amsmath}
\usepackage{amsthm}
\usepackage{braket}
\usepackage{mathdots}

\usepackage{makeidx}
\makeindex

\usepackage{soul}
\usepackage{xargs}
\newcommandx{\cmnote}[2][1=]{\linespread{1.0}\todo[linecolor=red,backgroundcolor=red!25,bordercolor=red,#1]{#2}}

\let\underline\ul

\allowdisplaybreaks[4]

 \index{} \index{} \index{} \index{} \index{} \index{} \index{} \index{}
\makeatletter

\newcommand{\ringplus}{\mathbin{\text{\@ringplus}}}

\newcommand{\@ringplus}{%
  \ooalign{\hidewidth\raise1.3ex\hbox{\tiny$\circ$}\hidewidth\cr$\m@th+$\cr}%
}

\newcommand{\ringminus}{\mathbin{\text{\@ringminus}}}

\newcommand{\@ringminus}{%
  \ooalign{\hidewidth\raise0.9ex\hbox{\tiny$\circ$}\hidewidth\cr$\m@th-$\cr}%
}
\makeatother
 \index{} \index{} \index{} \index{} \index{} \index{} \index{} \index{}

\DeclareFontFamily{U}{wncy}{}
\DeclareFontShape{U}{wncy}{m}{n}{<->wncyr10}{}
\DeclareSymbolFont{mcy}{U}{wncy}{m}{n}
\DeclareMathSymbol{\Sh}{\mathord}{mcy}{"58}

 \index{} \index{} \index{} \index{} \index{} \index{} \index{} \index{}
 \index{} \index{} \index{} \index{} \index{} \index{} \index{} \index{}

 \index{} \index{} \index{} \index{} \index{} \index{} \index{} \index{}
 \index{} \index{} \index{} \index{} \index{} \index{} \index{} \index{}
\usepackage{graphicx}

\usepackage{graphicx}
\usepackage{multirow}
\usepackage{hhline}

\xyoption{color}
\xyoption{line}

\newcommandx*\bsbal[3][1=black, 3=->]{\ar @[#1]@{#3} [#2,0] \qw}

\newcommandx*\varbs[5][1=black, 3=\theta,4=0.5,5=->]{\ar @[#1]@{#5}^(#4){#3} [#2,0] \qw}

\newcommandx*\lblline[3][3=0.5]{\ar @{-}^(#3){#1} [#2,0]}
\newcommandx*\ctrlg[3][3=0.5]{ \raisebox{-3pt}{$\bullet$}  \ar @{-}^(#3){#1} [#2,0] \qw }
\newcommandx*\ctrlog[2]{\controlo \ar @{-}^{#1} [#2,0] \qw}
\newcommandx*\ctrlodash[1]{\controlo \ar @{-} [#1,0] \ar @[black]@{.} [0,-1]}